# Photosensitive PEEK Ink Enables Digital Light Processing 3D Printed High-performance Small Architected-Plastics

Ze ZHANG[a, §], Kewei SONG[a, §], Rongyi ZHUANG[a], Jianxian HE[a], Yi YANG[b], Yifan PAN[a], Takeshi MINO[c], Kayo HIROSE[d,*] Shinjiro UMEZU[a,e,*]

[a]*Graduate School of Creative Science and Engineering, Department of Modern Mechanical Engineering, Waseda University, 3-4-1 Okubo, Shinjuku-ku, Tokyo 169-8555, Japan.*

[b]*Graduate School of Advanced Science and Engineering, Department of Integrative Bioscience and Biomedical Engineering, Waseda University, 3-4-1 Okubo, Shinjuku-ku, Tokyo 169-8555, Japan.*

[c]*Kagami Memorial Research Institute for Materials Science and Technology, 2-8-26 Nishiwaseda, Shinjuku-ku, Tokyo, 169-0051, Japan*

[d]*Anesthesiology and Pain Relief Center, The University of Tokyo Hospital, 7-3-1 Hongo, Bunkyo-ku, Tokyo 113-8655, Japan.*

[e]*Department of Modern Mechanical Engineering, Waseda University, 3-4-1 Okubo, Shinjuku-ku, Tokyo 169-8555, Japan.*

§: *The authors contributed equally.*

**Corresponding authors:**

*Shinjiro UMEZU, Ph.D., Professor, E-mail: umeshin@waseda.jp*

*Kayo HIROSE, Ph.D., E-mail: hirosek-ane@h.u-tokyo.ac.jp*

**Abstract**

Polyetheretherketone (PEEK), as a semi-crystalline high-performance engineering plastic, has demonstrated good application prospects since its introduction. The ability of PEEK to be fabricated in complex architecture is a major limitation due to the inherent shortcomings of material extrusion 3D printing technology in terms of low resolution, low surface quality, and interlayer bonding. We propose a novel PEEK ink processing process based on digital light processing (DLP) 3D printing, which is based on high solid content PEEK ink to achieve green bodies with high accuracy, and one-step sintering to enhance the crystallinity of PEEK. We have investigated the processing mechanism of this process and constructed perfect process parameters in terms of mouldability, printing accuracy, material thermal properties, and PEEK crystallinity. Furthermore, the material and architecture performance of the proposed process was evaluated in terms of comprehensive thermal performance (including heat resistance of the substrate, thermal stability, surface energy after heat treatment, and coefficient of static friction and coefficient of kinetic friction), mechanical performance, and corrosion resistance (20 wt% hydrochloric acid, 20 wt% sodium hydroxide, 99 wt% acetone, and 99.5 wt% chloroform). The process is a bold extension of PEEK processing methods to utilize the properties of PEEK in more flexible and efficient applications.

# 1. Introduction

Polyetheretherketone (PEEK) is a super-engineered plastic with a wide range of unique properties including high mechanical strength[1], good high-temperature resistance[2], excellent electrical properties, chemical stability[3], and biocompatibility. Due to these properties, PEEK has a wide range of applications in various fields such as automotive[4], aerospace[5-8], electronics industry[9], and biomedical[10, 11]. As a result, various molding processes for PEEK are constantly being developed to take advantage of its material benefits in a variety of important application scenarios[12-15].

Current PEEK molding processes mainly include injection molding[16], extrusion molding, compression molding, and computerized numerical control (CNC) machining methods for large size architecture processing[17, 18]. Although the injection molding process can give full play to the advantages of the mechanical properties of PEEK material, the long design cycle and high cost of mold development for complex architectures directly limit the flexibility of PEEK material application. Extrusion and compression molding, on the other hand, are more suited to large-scale production of simple architectures such as pipe and sheet. Such monolithic architecture require secondary processing to finalize the material afterward. CNC machining has the same problems when machining other raw materials, which are, high cost of high precision machining, complex spatial architecture that depend on process design and still have machining limitations.

With the development of additive manufacturing technology[7, 19-22], material extrusion additive manufacturing, selective laser melting (SLM), and selective laser sintering (SLS) methods are gradually being developed and applied to the machining of PEEK materials[11, 23-27]. High-temperature laser sintering of PEEK has been reported to approach the mechanical properties of injection-molded PEEK blends, a breakthrough from the drawbacks associated with previous SLS sintering[23]. However, it is important to mention that the anisotropy of the architectures of this process has not been investigated in depth and still has potential critical flaws. Typical of material extrusion, the fused deposition modeling (FDM) 3D printing method, where the raw material for processing is wire, brings significant savings in raw materials[28]. However, the processing strength of FDM is more than 30% lower than that of traditional processing methods, and most critically, the mechanical properties in the laminated direction are much lower than those in the vertical direction[29, 30]. In addition, the accuracy of FDM printing is often about 0.1mm, the surface quality of the architectures is poor, which cannot meet the requirements of precision machining, and face the problem of shrinkage in the cooling process in large-size machining, which makes the reliability of printed architectures questionable.

This study presents a new process for photosensitive PEEK Ink based on digital light processing (DLP) 3D printing technology and single-step monolithic sintering method. PEEK micron powders and photosensitive resins were mixed together in a planetary ball mill, without using dispersants, to create a PEEK ink that has a maximum solid content of 60 wt%. Green bodies were prepared layer by layer by a DLP 3D printer projecting a 405nm UV light pattern, and finally, the green bodies were sintered in a muffle furnace. In this paper, the principle and process parameters of the proposed PEEK ink printing process are studied and analyzed. Also, the thermal performance, mechanical properties, and chemical corrosion of the formed architectures are investigated. The results show that the proposed process can take full advantage of the PEEK material performance while achieves high-precision, flexible and easy-to-deploy architecture processing. Our research is expected to broaden the application prospects of PEEK materials especially where structural complexity and high resolution are required.

## 2. Result and Discussion

### 2.1 DLP 3D printing process

The PEEK ink printing process based on DLP 3D printing proposed in this study mainly consists of PEEK ink preparation, DLP 3D printing, and single-step finishing sintering. The core starting point is to achieve flexible preparation of PEEK ink through DLP 3D printing and to ensure and improve the printing accuracy as much as possible. The complete process is shown in **Fig. 1 a)**.

As a rapid prototyping technology, the high precision and resolution of light-cured 3D printing are directly affected by the material, as the design and preparation of PEEK inks are the basis for realizing the process. As a material that is not photosensitized by photosensitive groups, PEEK powder with an average particle size of 9 um was used as the raw material in this study. Based on the mouldability, solid loading capacity, and inherent properties of UV resins, 1,6-Hexanediol diacrylate (HDDA), Trimethylolpropane triacrylate (TMPTA), and Diethyl phthalate (DEP) were used to construct the crosslinked network in this study, in which Diphenyl(2,4,6-trimethylbenzoyl) phosphine oxide (TPO) was used as a photoinitiator. The resin system has good viscosity characteristics (as shown in **Fig. 2 b)**) which can effectively achieve high solid content powder loading. Based on the density and chemical stability of PEEK, 60 wt% solids PEEK inks were prepared by a planetary ball mill without the use of a dispersant (PEEK inks are referred to as 60 wt% PEEK inks when not otherwise specified in this paper), as shown in **Fig. 1b) I)**, the main components of the ink are in a mixed state.

The prepared PEEK ink is well-suited for the DLP 3D printing process. Process tests show that, under reasonable parameters, its printing accuracy surpasses the photosensitive resin system, reaches the theoretical resolution of the DMD in the commercial DLP 3D printer used, and effectively constructs complex aggregated architecture. This process is shown in **Fig. 1 b) II) and III)**, where the photosensitive polymer completes the cross-linking reaction under 405 nm UV light irradiation, and the encapsulated PEEK particles are transformed from a solid-liquid mixing state to a fully cured state. Based on the proposed process, we prepared the architecture of four lattices (Gylroid (**Fig. 1 c) I)**), Schwarz (**Fig. 1 c) II)**), Dode, and Trunc-Octa.), as well as a complex and beautiful model of the Eiffel Tower (shown in **Fig. 1. c) V)**. As shown in **Fig. 1. c) V)**, we compared the compression strength of different lattices printed with commercial resin and PEEK ink (the lattice arrangement is 4*4*4mm, please refer to the ***Supplementary Material*** for the compression strength measurement method), and the architectures prepared with PEEK ink can significantly improve the compression strength, which means that the process can provide a good protective shell for unmanned aerial vehicle (UAV), as shown in **Fig. 1. c) III)**. In an architecture with a porosity of more than 87%, in which the lattice is Dode, the plant leaves can carry its weight (shown as **Fig. 1. c) VI)**). In addition, the process can realize the high porosity architecture printing of PEEK ink material, and the porosity is better than that of the commercial resin under different lattice architectures (see ***Supplementary Material*** for commercial resin parameters). The proposed process well overcomes the limitations of PEEK machining under injection machining and machining processes.

As a semi-crystalline polymer, heat treatment can directly affect the mechanical properties, thermal properties, wear resistance, and other key properties of PEEK. Distinct from the regional thermal processing method such as SLM, the layer-by-layer curing in the light curing stage is performed at room temperature most of time, resulting in uniform thermal stress. A single-step burn in a muffle furnace achieves PEEK crystallization with a reasonable sintering curve under the part, to prevents deformation while removing internal stresses, ensuring good mechanical properties of the architectures. Sintering was completed under a nitrogen atmosphere and held near the $T_g$ temperature and melting point of PEEK, respectively, shown as **Fig. 1 b) IV)-VI)**, and the overall properties of the architectures were further enhanced by the crystallization of PEEK, although the photosensitive resin undergoes degradation in the process. Shown as **Fig. 1 c) IV)**, the lightweight and high-strength material can be applied to the architectures of the UAV and protective armor, combined with the high temperature resistance and chemical corrosion resistance characteristics of PEEK, provides a good guarantee for the UAV to perform special search tasks in scenarios such as chemical and fire accidents. **Fig. 1 c) V)** shows a highly porous lightweight architecture that can carry the weight of a plant leaf.

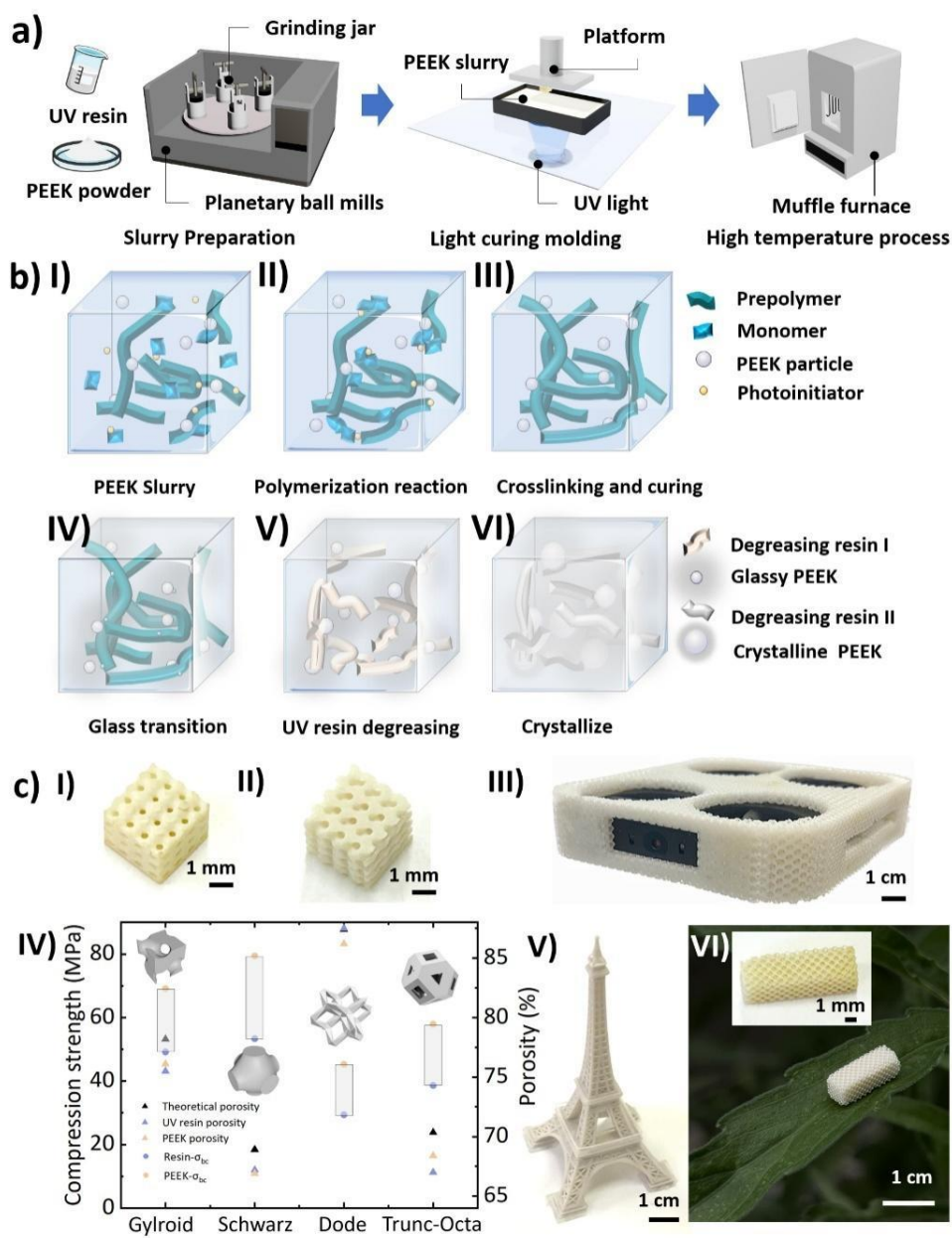

**Figure 1. The framework for the study of DLP 3D printing processes for PEEK ink.** a) PEEK ink machining process based on DLP 3D printing, including ink preparation, DLP 3D printing and heat treatment processes, b) Network

cross-linking processes in PEEK ink UV irradiation (I)-III)) and molecular changes under heat treatment (IV)-VI)). c) Complex structures prepared by the proposed PEEK ink 3D printing process. I) and II) are triply periodic minimal surface-based topologies (Schwarz and Gylroid) with lattice units of 1 mm, respectively. III) is a lightweight protective shell structure for quadrotor application, the lattice is made of Trunc-Octa, and the unit lattice size is 3*3*3mm. IV) Compressive strength and porosity of porous structures prepared by commercial resins and the proposed PEEK ink 3D printing process with lattice type Gylroid, Schwarz, Dode, and Trunc-Octa. V) is Eiffel Tower model after sintering. VI) is an architecture with a porosity of more than 87%, and the lattice is Dode, the plant leaves can carry its weight.

There are many methods to add fillers to liquid photosensitive resins for part modification[31], it is usually necessary to avoid the influence of excessive fillers on the main properties of the substrate. However, the photosensitive PEEK ink DLP 3D printing processing process proposed in this study, in which the core material properties are achieved through PEEK particles, is similar to the processing of ceramics. Therefore, the solid content of the lifting material needs to be increased as much as possible while being able to balance mouldability and machining accuracy. In this study, the bifunctional monomer HDDA is used to provide mechanical strength and toughness to the substrate. The trifunctional monomer TMPTA enhances the density of the crosslinked network, imparting heat resistance, high hardness, and wear resistance, to prevent excessive rigidity and brittleness of the cured material. DEP is used as a plasticizer to improve the flexibility of the material, and to reduce the glass transition temperature of the resin to match the window for the PEEK sintering process. In this photosensitive resin system, a planetary ball mill was used to achieve physical dispersion of the PEEK particles, achieving a maximum mouldable solid content of 60 wt%, material delaminates after 6 hours of resting, area stabilizes after 12 hours (Refer to **Supplementary materials** for results). The up-forming DLP 3D printing process relies on the fluidity of the material, with the printing platform moving up and down in the direction of lamination (usually the Z-axis), which is naturally replenished layer by layer, so that the material rheology should not be too high in the range of shear rates between 100-500 s$^{-1}$. The rheology of PEEK particles with a solids content of 0 wt%, 20 wt%, 40 wt% and 60 wt% is shown in **Fig. 2 b)**. The rheology at 60 wt% exceeds 2000 mPa·s, which is close to the printable limit.

The printing error of PEEK-60wt% ink in the X, Y, Z directions under different exposure times was analyzed (please refer to the **Supplementary Material** for the error evaluation method), shown as **Fig. 2a)**. Using the photosensitive resin used as a control group, the error of PEEK ink is significantly smaller than that of photosensitive resin under a single-layer exposure time of less than 12 s, and the error of the resin shows a decreasing trend with the increase of exposure time. The PEEK in the ink affects the overall light transmittance and absorbs part of the UV light, so the overall printing accuracy in the X, Y directions are higher, but the Z-direction error tends to increase with the increase of the exposure time, and the solid particles in the tight cross-linking network leads to an increase in the expansion in the Z-direction.

The thermogravimetric analysis (TGA) results of the cured photosensitive resin as well as the PEEK raw material powder are shown in **Fig. 2c)**, where PEEK demonstrates consistent thermal stability, but the cured resin starts to degrade near 250 ℃ and the degradation is accelerated at 400 ℃, so the overall sintering temperature is initially controlled at the 280-340 ℃ stage. The sintering process curve is shown in **Fig. 2 d)**, the Tg temperature and the maximum temperature of PEEK were reached at an overall heating rate of 10 °C/min, respectively, and were held for 20 and 40 min, respectively, and then cooled with the furnace in order to complete the sintering process of the experimental samples (for sample dimensions and design criteria, please refer to the ***Supplementary Material***). The sintering process was carried out with the whole in a nitrogen atmosphere to avoid the oxidation of the samples. The PEEK particles provide good thermal stability for the architecture as a whole, and under the sintering curves at different maximum temperatures, the deformation of the photosensitive resin parts increases significantly, and the PEEK parts remain essentially stable, as shown in **Fig. 2 e)**. The green body crystallinity of PEEK, detected by X-ray diffraction (XRD), is about 15%, which increases with isothermal temperature, and the highest crystallinity of about 37% is obtained at 340 °C, as shown in **Fig. 2 f)**.

The thermal stability of PEEK materials has been extensively studied[32], but the influence of the photosensitive resin system by the thermal sintering process needs to be further explored, and the Fourier-transform infrared spectroscopy (FTIR) of PEEK and resin are shown in **Fig. 2 g) I) and II)**, respectively [33]. Alkyl chains (C-H) in the resin at 3000-2800 $cm^{-1}$, mainly from HDDA and TMPTA, telescopically vibrate at 280-300 °C, and the vibration is weakened at 320 °C, probably due to partial breakage of the alkyl chains at elevated temperatures, and decompose significantly at 340 °C. The characteristic carbonyl (C=O) peaks at 1700-1600 $cm^{-1}$ are from HDDA, TMPTA, and DEP, which follow the same trend under heat as the alkyl chains, with substantial decomposition at 340 degrees Celsius. The ether bond (C-O-C) at 1300-1000 $cm^{-1}$ is relatively stable at high temperatures until it is partially broken at 340 °C. One of the aromatic ring structures breaks completely at the high temperature stage. In contrast, when PEEK particles are present, the chemical structure of the material remains stable at two temperatures, 280 °C and 300 °C, with no significant changes in the major chemical bonds. Upon increasing the temperature to 320 °C, the spectra show some slight changes, but the major chemical bonds (including C=C and C-O-C) remain. At 340 °C, changes in the characteristic peaks indicate that the aromatic ring and ether bonds are beginning to be affected.

It is worth noting that UV irradiation has some effect on the molecular chain integrity of PEEK during light-cured printing, which is directly reflected in the change in surface energy[5, 34]. In **Fig. 2 h)** the water contact angles of the photosensitive resin and PEEK architectures are shown separately, where the water contact angle of PEEK is kept at 70 °C, so it is necessary to limit the time of UV irradiation of its single layer in order to avoid the aging of PEEK.

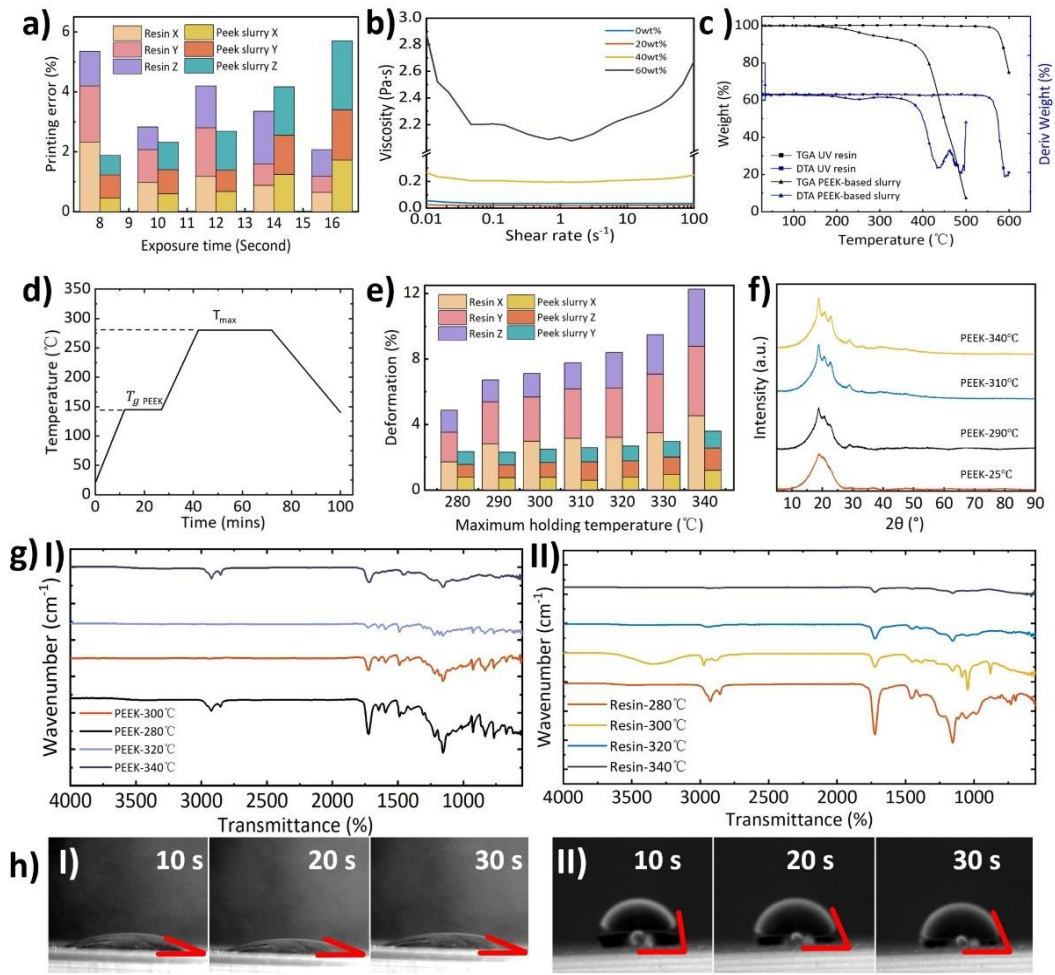

**Figure 2. Key parameters of the printing process.** a) The effect of different exposure times on the printing accuracy error of the UV resin used in the X, Y, and Z directions, b) Viscosity of PEEK inks with solid content, c) Thermal stability analysis of the used UV resins and PEEK powder at 0-600 °C, d) Green bodies one-step sintering curve, $T_{max}$ were 280 °C, 290 °C, 300 °C, 310 °C, 320 °C, 330 °C, and 340 °C, respectively, e) Results of the effect of sintering at different maximum holding temperatures on the shrinkage of the part in the X,Y,Z direction, f) **XRD** results for crystallinity analysis under different sintering profiles, g) FT-IR results of the materials under UV curing and sintering, I) the heat treatment reaction results of the PEEK inks under different sintering profiles, II) the curing cross-linking process of the used UV resin at room temperature and the heat treatment results under different sintering profiles, h) Water contact angle of the material after UV curing, I) UV resin used, II) PEEK inks.

## 2.2 Heat treatment performance

PEEK tends to be applied in high temperature working environments[35, 36], so the change in its properties after heat treatment needs to be further verified. The microscopic behavior of the resin material at different temperatures was obtained by scanning electron microscope (SEM), as shown in **Fig. 3 a)**, **I)** shows the surface of the resin part without sintering, which remains intact, and the cracks in the part increase and increase in size with increase in heat treatment temperature. **Fig. b I)** shows the surface of the part after sintering of PEEK at 280 ℃, which remains intact, although the presence of PEEK particles can be observed, **II) and III)** are artificial sections of untreated and isothermally held PEEK parts at 340 ℃, respectively, and the average particle size of the unfired PEEK is lower than that of the part after high-temperature sintering, with a rougher surface of the particles, **VI)** is a magnified view of the sintered particles, where the PEEK can be observed to be fused and connected to each other.

**Fig. 3. c)** compares the thermogravimetric analysis of the resin substrate and PEEK after high temperature treatment; the weight loss of PEEK is significantly lower than that of the resin, and by removing the variations under the percentage of PEEK, the overall mass loss is still lower than that of the resin parts, which indicates an improvement in the overall thermal stability of the material. **Fig. 3 d)** shows the results of differential scanning calorimetry (DSC) of the resin and PEEK in the primary heating, primary cooling and secondary heating. PEEK undergoes crystal reorganization in the primary heating and has a large exothermic amount during the cooling process, which means that it still possesses the typical characteristics of PEEK after the process proposed in this paper. Resins, on the other hand, emit less heat, and since they are non-crystalline materials, changes after high temperature treatment are mainly due to decomposition.

The thermal treatment may again change the surface energy of the material, where the resin remains consistently hydrophilic and the PEEK surface exhibits a tendency for the water contact angle to decrease with increasing isothermal temperature at different temperatures, with an increase in the surface energy potentially indicating that the cells are more readily attached to the surface of the material.

The surface architecture of the part changes after sintering, and with it, its roughness. We measured the coefficients of kinetic and static friction of resin and PEEK parts on 100 grit sandpaper (The measuring device is shown in **Fig. 3 g) I)**). As shown in **Fig. 3 g) II)**, the coefficients of kinetic and static friction of the resin tended to increase as the isothermal temperature increased, whereas the coefficients of kinetic and static friction of the PEEK showed a decreasing tendency and reached a minimum at 320 °C. The values are shown in **Table 1** (see *Supplementary Material* for the methodology of the measurements).

Table 1

**Experiments on dynamic and static friction with different sintering profiles.**

| Friction coefficient | | Temperature | | | | | | | |
|---|---|---|---|---|---|---|---|---|---|
| | | 25 °C | 280 °C | 290 °C | 300 °C | 310 °C | 320 °C | 330 °C | 340 °C |
| PEEK | Static | 0.842 | 0.830 | 0.807 | 0.764 | 0.764 | 0.738 | 0.744 | 0.749 |
| | Kinetic | 0.677 | 0.709 | 0.705 | 0.699 | 0.699 | 0.662 | 0.670 | 0.742 |
| Resin | Static | 0.854 | 0.936 | 0.932 | 0.929 | 0.91 | 0.886 | 0.910 | 0.929 |
| | Kinetic | 0.776 | 0.763 | 0.780 | 0.789 | 0.817 | 0.824 | 0.841 | 0.854 |

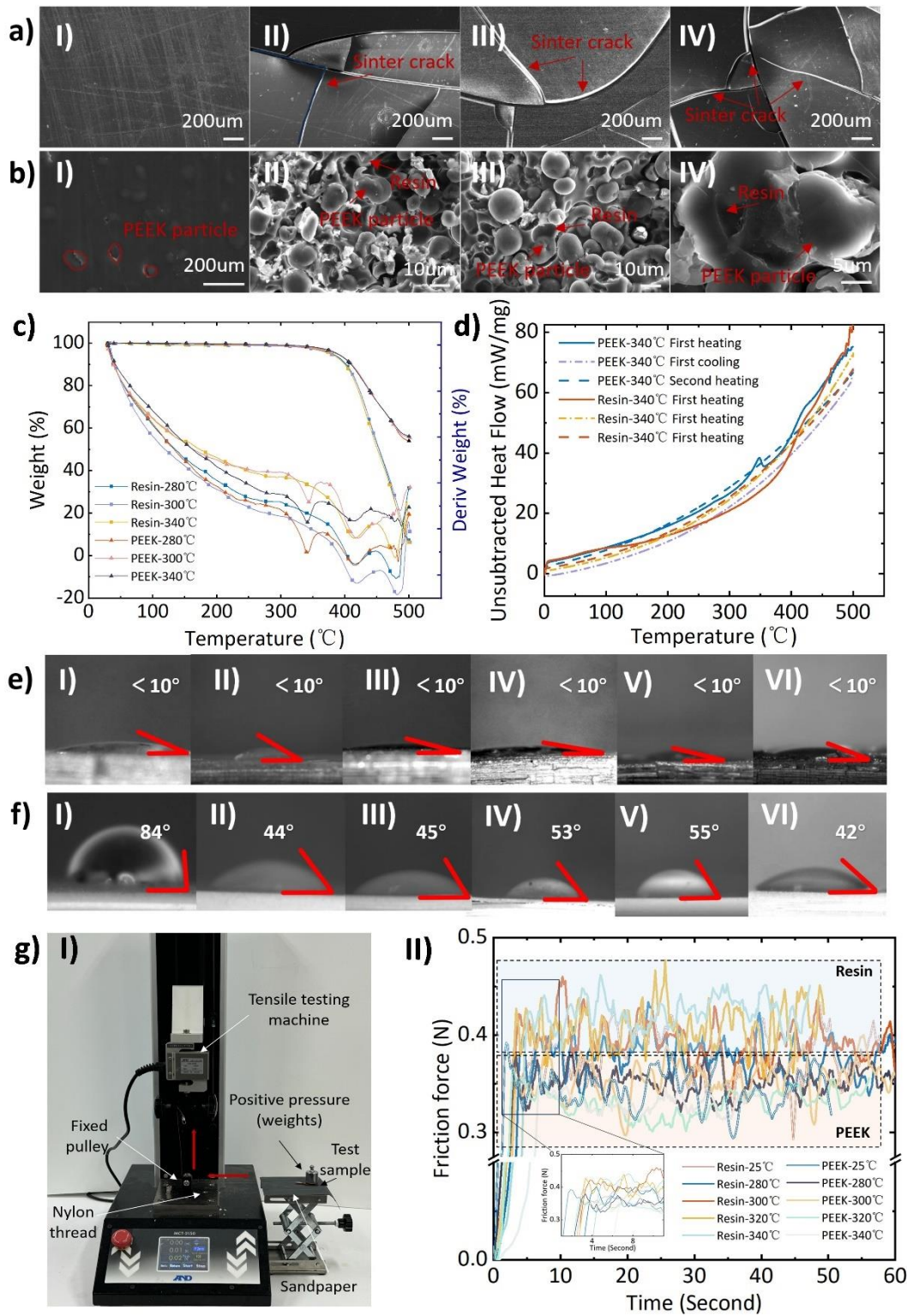

**Figure 3. Part stability, glass transition temperature, water contact angle, and coefficient of friction properties under heat treatment processes.** a) Microstructure of resin parts under different heat treatment conditions, from I) to IV) are the SEM images of the part surface without heat treatment (25 °C), 280 °C, 300 °C and 340 °C, respectively, b) Microstructure of PEEK parts under different heat treatment conditions, from I) to IV) are the SEM images of the part surface without heat treatment (25 °C), 280 °C, 340 °C and 340 °C, respectively, c) Results of stability analysis after heat treatment based

on TG analysis, d) DSC analysis, performance analysis after elimination of thermal history, e) Variation of water contact angle at different heat treatment profiles for the UV resins used, f) Variation of water contact angle of PEEK inks under different heat treatment profiles, g) Measurement of coefficient of friction, I) Measuring devices and methods, II) Stick-slip behavior of PEEK and resins on 100 grit sandpaper for different process conditions.

## 2.3 Mechanical properties

After printing the part, as high performance engineering materials, both PEEK and resin influence the mechanical properties of the printed structure. Heat treatment enhances the crystallinity of PEEK which in turn affects the mechanical properties, however, sintering of the substrate, which is a thermosetting material and is accompanied by thermal degradation, leads to side effects on the mechanical properties. **Fig. 4 a)** illustrates the mechanical performance of different PEEK solids at 280 °C isothermal temperature, while tensile, bending, and compression tests were performed using the resin without heat treatment and the resin parts under the same thermal conditions as a control group, for test standards and experimental parts and conditions, please refer to the ***Supplementary Materials***. In this case, the plastic deformation phase can hardly be observed in the tensile stress-strain curve, so that the material as a whole maintains the characteristics of a brittle material. The solid content of PEEK shows an increase in strength with increasing content in both tensile and bending, with similar results within the compressive strength gradient.

The effect of different isothermal temperatures on the mechanical properties is illustrated in **Fig. 4 b)**, where the dashed segments indicate the results for green bodies that have not been heat-treated, in **I)**, the tensile strength and the maximum strain basically increase in parallel with the temperature, in **II)**, it can be observed that the bending strength increases with the temperature but the maximum strain shows a decreasing tendency, and in compression tests, the highest strength is found at 340 °C but the strain is in an intermediate position. It is clear that the increase in isothermal temperature and solid content effectively improves the mechanical properties of PEEK parts.

The tensile sections of the resin and PEEK specimens without heat treating are shown in **Fig. 4 c) and d)**, the resin section consists of slip marks and is clean, and the PEEK section exists in a state of particle tearing, where **III)** and **IV)** are the results of energy dispersive spectrometer (EDS) analyses for the corresponding carbon and oxygen elements, respectively. **Fig. 4 e)** shows the tensile fracture interface diagrams of PEEK parts at different isothermal temperatures from **I)** to **IV)** 280 °C, 300 °C, 320 °C, and 340 °C. It can be noted that the strength is provided by the forces between the resin matrix, between the PEEK particles, and between the PEEK and the resin when the specimen is damaged. The cross-section of the PEEK between the matrices is clear and clean, so it is presumed that the forces do not increase significantly as the temperature increases. On the other hand, the resin decreases in strength with increasing temperature and is found to be partially carbonized when removed during the muffle furnace sintering test. resulting in very brittle during the sintering test and thus not able to withstand the clamping force of the tensile testing machine fixture clamping force and breakage (As shown in **Fig. S2**). Therefore, the main contribution to the mechanical property enhancement brought about by heat treatment comes from the mechanical property enhancement of PEEK after high temperature heat treatment.

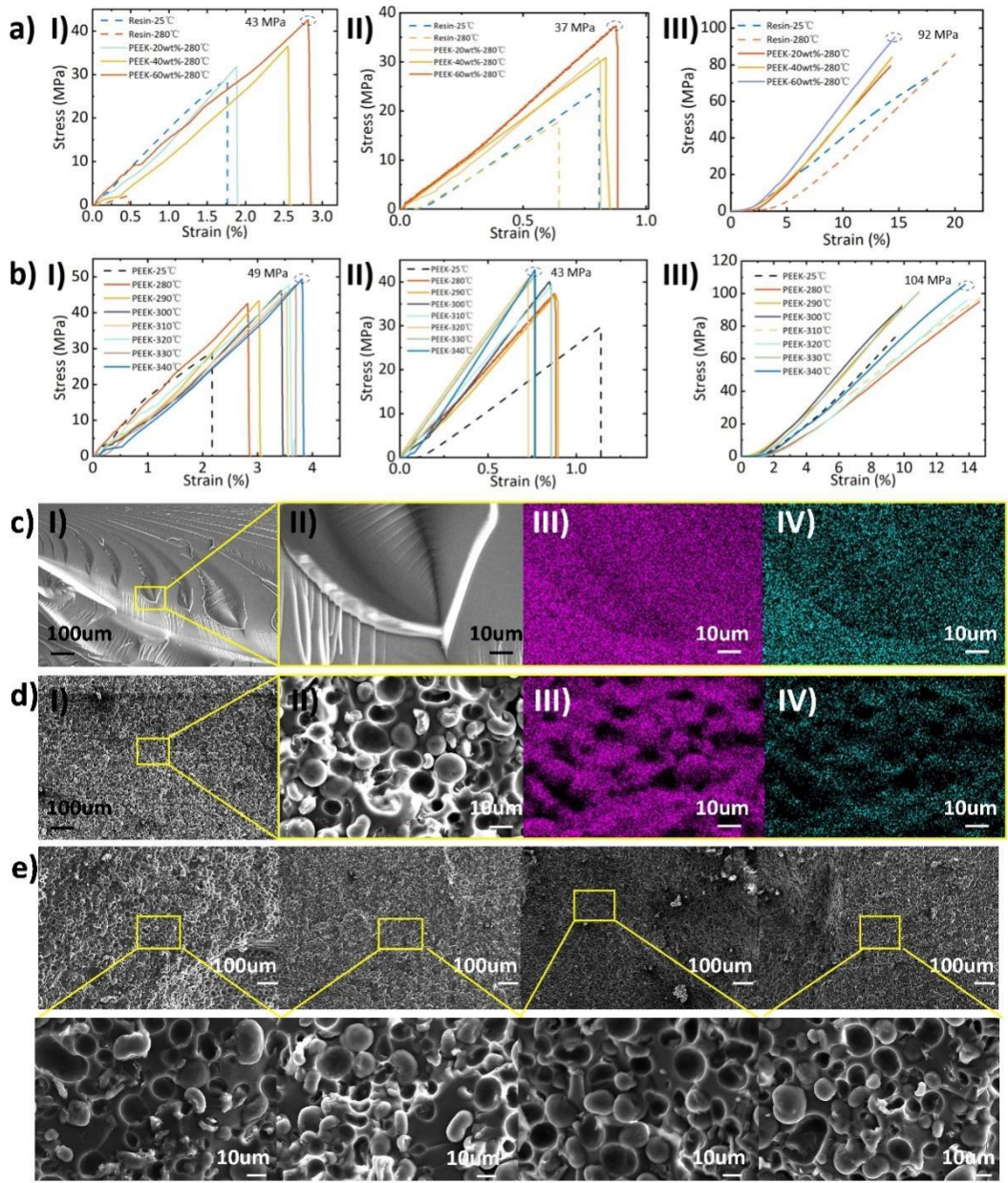

**Figure 4. Mechanical property test results under different process parameters.** a) Mechanical properties of the photosensitive resin used and PEEK at concentration gradients of 20 wt%, 40 wt%, 60 wt%, I)-III) are stress-strain curves for tension, bending and compression, respectively, b) Mechanical properties after different heat treatment processes at a PEEK concentration of 60 wt% with maximum holding temperatures of 280 °C, 290 °C, 300 °C, 310 °C, 320 °C, 330 °C and 340 °C, respectively, I)-III) are stress-strain curves for tension, bending and compression, respectively, c) Tensile fracture microstructure of photosensitive resins without heat treatment, I)-IV) are scanning electron microscope (SEM) images and energy dispersive spectrometer (EDS) of carbon and oxygen elements, respectively. d) Tensile fracture microstructure of 60 wt% PEEK without heat treatment, I)-IV) are SEM images and EDS of carbon and oxygen elements, respectively. e) SEM images of tensile fracture sections under different heat treatment processes, the maximum heat treatment temperatures from I) to IV) were compared to 280 °C, 300 °C, 320 °C and 340 °C.

**Table 2**

**Comparison of mechanical properties of PEEK processed by different additive manufacturing.**

| Process | Mechanical properties | | | Material | Ref. |
|---|---|---|---|---|---|
| | Tensile strength (MPa) | Bending strength (MPa) | Compression strength (MPa) | | |
| FDM | 89.23 | 106.45 | 103.74 | Thermax PEEK | [37] |
| | 84 | - | 135 | 450G | [38] |
| | 56.6 | 56.1 | 60.9 | 550G | [39] |
| SLS | 88.7 ± 1.5 | 123.0 ± 2.5 | 184 ± 15 | EOS PEEK HP3 | [23] |
| | 97.7 ± 0.85 | - | - | EOS PEEK HP3 | [40] |
| Proposed DLP | 49 | 43 | 104 | PEEK T010-950M | - |

## 2.4 Corrosion resistance

PEEK has good chemical stability and exhibits corrosion resistance in a wide range of solutions[41]. We tested and observed unsintered (25 °C), 280 °C-sintered and 340 °C-sintered PEEK and resin architectures for corrosion resistance testing. Chemicals of choice are 20 wt% hydrochloric acid (HCl), 20 wt% sodium hydroxide (NaOH), chloroform (≥99.5 wt%), and acetone (99.5 wt%). Samples were placed in sealed solvent bottles for immersion, weighed, and surface observations were made at time intervals of 20 mins, 40 mins, 60 mins, 120 mins, and 720 mins, refer to ***Supplementary Materials*** for reagent bottle records.

The test results are shown in **Fig. 5**, where the left side of each group shows the image record of the initial state of the samples, the right side shows the image record of the surface of the samples at 720 mins, and the middle shows the curve of weight change with time. The maximum mass loss of PEEK in NaOH, HCl, and acetone were all within 5 wt%, and reached 12 wt% in chloroform, with mass loss decreasing with increasing isothermal temperature and decreasing corrosion resistance, while the overall PEEK corrosion resistance was higher than that exhibited by the resin under all types of conditions. The trend of mass loss is regionally stable over a long period of observation and the chemical resistance of the proposed process is good.

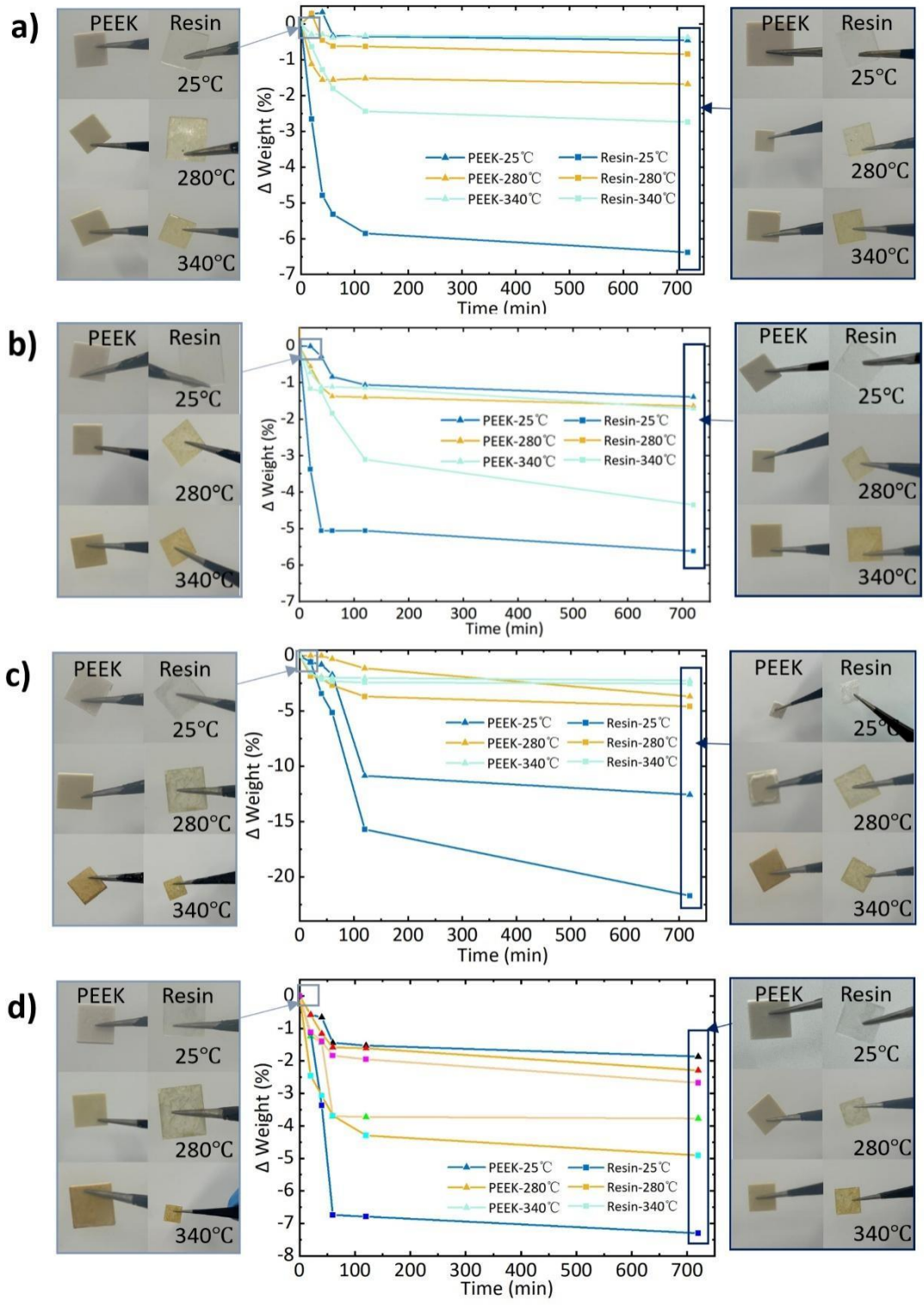

**Figure 5. Chemical corrosion test.** a) 20 wt% sodium hydroxide (NaOH) solution, b) 20 wt % hydrochloric acid (HCl) solution, c) Chloroform, d) Acetone.

## 3. Conclusion

We propose a photosensitive PEEK ink processing based on DLP 3D printing, enhance the crystallinity of PEEK by one-step sintering in order to achieve a high-precision machining process for complex spatial architectures of PEEK materials. Among them, the processing principle, process parameter selection and architectures properties were investigated. It is shown that based on the proposed resin system, the weight ratio of PEEK particles in photosensitive resin is 60 wt%, and under the DLP printer with X, Y-axis optical resolution of 51 um, the printing accuracy can be basically the same as that of the pure photosensitive resin and the commercial resin by using the experimentally verified process parameters, which can achieve the ability of complex architecture printing of the photosensitive PEEK ink. Also, the compressive strength and actual porosity of its various lattices are demonstrated, which have higher strength and closer to the theoretical porosity in comparison with commercial photosensitive resins. The study of one-step crystallization heat treatment process shows that the mechanical properties are directly related to the PEEK solid content and the highest isothermal temperature of sintering, and under the isothermal condition of 340°C, the PEEK ink parts reach the tensile strength of 48 MPa, the bending strength of 43 MPa, and the compression strength of 105 MPa respectively under the standard mechanical test; at the same time, the proposed PEEK ink printing process improves the overall material stability, which was verified by DSC and TGA analyses. In addition, UV irradiation provided an in-situ surface treatment, and the surface energy of PEEK was significantly enhanced and remained high after heat treatment, potentially improving the cell adhesion ability. Moreover, the PEEK surface properties remained relatively stable in chemical resistance tests with 20 wt% NaOH, 20 wt% HCl, chloroform and acetone. The proposed process for PEEK inks can effectively take advantage of PEEK materials, which can be extended to photosensitive resins with different properties to adapt to different working scenarios, including mechanical engineering, aerospace, medical implantation, therefore has foreseeable potential.

## *4. Methods and Experiments*

### 4.1 Materials

Polyetheretherketone (PEEK), product code T010-950M with an average diameter (D50) of 14 μm, was purchased from T&T Industry Group Ltd., China. The photosensitive resin monomers include 1,6-Hexanediol diacrylate (HDODA), Trimethylolpropane triacrylate (TMPTA) and Diethyl phthalate (DEP) obtained from Sigma-Aldrich Co., Ltd. And Diphenyl(2,4,6-trimethylbenzoyl) phosphine oxide (TPO) used as a photoinitiator. Hydrochloric acid (20 wt%) and acetone (Guaranteed reagent, 99.5 wt%) was obtained from FUJIFILM Wako Pure Chemical Co., Ltd. Sodium hydroxide and chloroform (≥99.5 wt%) was obtained from Sigma-Aldrich Co., Ltd. Please refer to the Supplementary Materials for the specific material properties.

### 4.2 Ink preparation

Photosensitive PEEK ink preparation includes two parts: photosensitive resin preparation and PEEK ink. The photosensitive resin was mixed with HDDA, TMPTA, and DEP in the ratio of 6:3:1 by weight, and 3 wt% of the total mass of TPO was added to the photosensitive resin, which was stirred by a magnetic rotor at room temperature (25°C) for 30 min under the condition of shading and at a speed of 100 r/min.

PEEK ink mixing is done with planetary ball mills (JZ-CXQ-400A, Zhongyi Instrument Co., Ltd., China), wherein, the grinding jar and the grinding ball material are corundum. According to the preparation of the completed photosensitive resin: PEEK powder quality of 4:6 weighing PEEK powder, respectively, in accordance with the proportion of 50 wt%, 30 wt%, 10 wt%, and 10 wt% of the corresponding photosensitive resin and PEEK powder to add to the grinding jar, in turn, grinding 3 hours, 2 hours, 1 hour, 1 hour until all the materials to add the completion of the grinding again at room temperature for 6 hours. After mixing, take out the ink in the vacuum condition heating it to 50 °C to keep for 20 mins to eliminate the air bubbles in the ink, and complete the ink preparation.

### 4.3 Additive manufacturing

PEEK-based green bodies were prepared using a DLP device with a UV wavelength of 405 nm (Photon D2, Shenzhen Anycubic Technology Co., Ltd., China). Firstly, the 3D digital model was completed in 3D modelling software and the model was sliced using Anycubic software. After setting the printing parameters, the PEEK ink was poured into the cassette to complete the part preparation. After printing, the parts were removed from the printing platform and cleaned with ethanol in an ultrasonic cleaner for 5 mins, and finally irradiated with green bodies at 405 nm for 30 s to complete the secondary curing, and then dried and used for sintering preparation. Please refer to the **Supplementary Materials** for the specific print parameter settings.

### 4.4 Sintering process

As shown in **Fig. 2 d)**, PEEK green bodies were ramped up at 10 °C /min and held at 145 °C and 280-340 °C (10 °C gradient temperature), respectively, to complete the PEEK crystallization process. In this case, the holding time of the standard block was 15 and 30 mins at 145 °C and maximum temperature, respectively. The complete sintering process was under $N_2$ atmosphere to avoid oxidization of the parts.

### 4.5 Characterizations

FTIR, XRD, DSC, TGA, SEM, EDS, and mechanical testing were performed in the study. The experiments were measured according to the consensus method, in which FT/IR-4200 (Japan Spectroscopy) equipment was used to perform FTIR; X-ray refraction device-Miniflex (Rigaku) was used to perform XRD; DSC-8500 (PerkinElmer, Inc, USA) equipment was used to perform DSC; TG-8120 thermogravimetric differential thermal analyzer (Rigaku) was used to perform TGA; JSM-5500 was used to perform SEM and EDS (Japan Electronics) equipment; mechanical testing experiments were performed by MCT-2150 bench-top tensile and compression testing machine (A&D). Other experimental tests include measurement of 3D printing accuracy, standard part density, density measurement by Archimedes' drainage method, dynamic and static friction coefficients and viscosity and rheology. Please refer to the **Supplementary Materials** for the specific measurement methods.

**Supporting Information**

● Peek material properties; 3D printing parameters set in this study；Measurement methods and experiments of 3D printing parts accuracy; Density measurement；Stability of the prepared peek；Measurement of static and dynamic friction coefficients；Rheology Test；Water contact angle measurement method； Corrosion resistance test； Mechanical properties testing process; Mechanical properties testing process of lattice；Porosity measurement method. (**.Word**)

**Acknowledgments:**

The authors thank Atsuko Nagataki Researcher and Kagami Memorial Research Institute for Materials Science and Technology, Waseda University for helping to experiments; Thank Advanced Research Infrastructure for Materials and Nanotechnology (ARIM); Thanks to Mr. Xu Zhaolun, who has graduated from Waseda University, for his early experimental support; Thanks to Mr. Chaolun XU for his help in experiments; This work was the result of using research equipment (G1023) shared in MEXT Project for promoting public utilization of advanced research infrastructure (Program for supporting construction of core facilities) Grant Number JPMXS0440500023; This work was supported by JSPS KAKENHI Grant Number JP 23H01382 and 23H01374.

**Credit authorship contribution statement:**

**Ze ZHANG:** Methodology, Investigation, Writing-draft, review & editing, Visualization, Data curation. **Kewei SONG:** Methodology, Investigation, Writing-draft, review & editing, Visualization. **Rongyi ZHUANG:** Visualization, Data curation. **Jianxian HE:** Visualization, Data curation. **Yi YANG:** Data curation. **Yifan PAN:** Visualization. **Takeshi MINO:** Data curation. **Kayo HIROSE:** Investigation, Writing - review & editing, Visualization, Supervision, Project administration, Funding acquisition. **Shinjiro UMEZU:** Conceptualization, Methodology, Investigation, Writing - review & editing, Visualization, Supervision, Project administration, Funding acquisition.

**Competing Interest:**

The authors report no declarations of interest.

**Data Availability Statement:**

The authors declare that the data supporting the findings of this study are available within the paper and its supplementary information files.